\documentstyle[graphicx]{mn}

\title{The identification of an optical counterpart to the
super-Eddington X-ray source, NGC 5204 X-1}

\author[T.\,P. Roberts et al. ]
{T.\,P. Roberts, M.\,R. Goad,
M.\,J. Ward, R.\,S. Warwick, P.\,T. O'Brien, P. Lira \and and
A.\,D.\,P. Hands\\
Department of Physics and Astronomy, University 
of Leicester, University Road, Leicester, LE1 7RH}
\date{}

\def\ro{{\it ROSAT~\/}}
\def\asca{{\it ASCA~\/}}
\def\ein{{\it EINSTEIN~\/}}

\def\chan{{\it CHANDRA~\/}}

\def\ergcms{{\rm ~erg~cm^{-2}~s^{-1}}}

\def\ergsec{{\rm ~erg~s^{-1}}}

\def\atpcm{{\rm ~atoms~cm^{-2}}}
\def\ctsec{{\rm ~count~s^{-1}}}

\def\H0{{\rm ~km~s^{-1}~Mpc^{-1}}}

\def\eg{{\it e.g.~\/}}

\def\la{\mathrel{\hbox{\rlap{\hbox{\lower4pt\hbox{$\sim$}}}{\raise2pt\hbox{$<$}}}}}
\def\ga{\mathrel{\hbox{\rlap{\hbox{\lower4pt\hbox{$\sim$}}}{\raise2pt\hbox{$>$}}}}}

\def\d25{D$_{25}$}
\def\nh{{$N_{H}$}}

\def\.25{0.25 keV\thinspace}

\begin{document}

\maketitle

\begin{abstract}
We report the identification of a possible optical counterpart to the
super-Eddington X-ray source NGC 5204 X-1.  New \chan data shows that
the X-ray source is point-like, with a luminosity of $5.2 \times
10^{39} \ergsec$ (0.5 - 8 keV).  It displays medium- and long-term
X-ray variability in observations spanning a period of 20 years.  The
accurate \chan position allows us to identify a blue optical continuum
source ($m_v = 19.7$) at the position of NGC 5204 X-1, using
newly-obtained optical data from the INTEGRAL instrument on the
William Herschel Telescope.  The X-ray and optical source properties
are consistent with the scenario in which we are observing the beamed
X-ray emission of a high-mass X-ray binary in NGC 5204, composed of an
O star with either a black hole or neutron star companion.
\end{abstract}


\section{Introduction}

{\it EINSTEIN\/} imaging observations were the first to reveal that
the X-ray emission of many spiral galaxies is dominated by a small
number of luminous (L$_{\rm X} \sim 10^{38-40} \rm ~erg~s^{-1}$)
discrete X-ray sources located outside the nucleus of the galaxy
(Fabbiano 1989 and references therein).  {\it ROSAT\/} and {\it
ASCA\/} observations have since confirmed the presence of one or more
luminous extra-nuclear sources in many individual nearby galaxies, for
example Dwingeloo 1 (Reynolds et al. 1997), Holmberg II (Zezas et
al. 1999), NGC 5194 (Marston et al. 1995) and NGC 4321 (Immler et
al. 1998).  Early results from the \chan observatory have extended
this theme, most notably with the resolution of a high luminosity
(L$_{\rm X} \sim 10^{40-41} \rm ~erg~s^{-1}$), variable X-ray source
located just off-nucleus in the nearby starburst galaxy M82 (Kaaret et
al. 2001; Matsumoto et al. 2001).  In a recent survey using archival
{\it ROSAT\/} HRI observations, Roberts \& Warwick (2000; hereafter
RW2000) catalogue 29 sources with L$_{\rm X} > 10^{39} \rm
~erg~s^{-1}$ in the 0.1 - 2.4 keV \ro band in a variety of nearby
galaxies.  Each of these sources is at least as luminous as a 10
M$_{\odot}$ black hole accreting at close to its Eddington limit;
however, the nature of these super-Eddington sources (SES\footnote{A
note on taxonomy; we refer to the class of L$_{\rm X} > 10^{39} \rm
~erg~s^{-1}$ extra-nuclear sources as SES throughout this letter.
This class of object is alternatively referred to as SLS, IXOs, or
ULXs in the literature, though the last acronym refers in particular
to objects suspected to be powered by accretion.}) remains unclear.
Known candidates for explaining the SES phenomena include recent
supernovae/very young supernova remnants (SNR), accretion onto a
compact object, or unresolved complexes of X-ray sources.

Supernovae exploding into a dense circumstellar medium can produce
very X-ray luminous nebulae in their early stages (\eg SN 1988Z with
L$_{\rm X} \sim 10^{41} \rm ~erg~s^{-1}$, Fabian \& Terlevich 1996),
and indeed several of the RW2000 SES sample can be identified with
recent supernovae, for example SN 1979C $\equiv$ NGC 4321 X-4 (Immler
et al. 1998) and SN 1986J $\equiv$ NGC 891 X-3 (Bregman \& Pildis
1992).

Alternatively, accretion processes may power much of the SES
population.  Evidence in favour of accretion comes from an {\it
ASCA\/} study of several SES (Makishima et al. 2000), in which the
spectra are shown to be well fit by multi-colour disc blackbody (MCD)
emission from an optically thick standard accretion disc around a
black hole.  The unusually high temperatures required to fit these
systems suggest that SES are analogous to the Galactic micro-quasars
GRO 1655-40 and GRS 1915+105, and may be powered by accretion onto
rapidly-rotating high stellar mass (few tens of M$_{\odot}$) black
holes.  In a similar vein, Colbert \& Mushotzky (1999) suggest that
some SES observed to be marginally offset from galactic nuclei (on
scales of several hundred parsecs) in {\it ROSAT\/} HRI observations
are not a type of low-luminosity AGN, but in fact constitute a new
class of ``intermediate'' mass accreting black holes having masses of
$10^2 - 10^4$ M$_{\odot}$.  The key to recognising SES as accreting
systems is, of course, via significant variability; indeed, some
systems are already known to display short-term variability over
timescales of $\sim 1000$ s (IC 342 X-1, Okada et al. 1998; Holmberg
II source, Zezas et al. 1999).  Also, recent {\it ASCA\/} studies have
demonstrated that long-term variability is present in several SES, and
is accompanied by transitions in the X-ray spectra of the objects
between ``hard'' and ``soft'' states (Mizuno et al. 2001; Kubota et
al. 2001).

Unfortunately, X-ray data from the pre-\chan era are severely
spatially limited and so moderately-compact complexes of X-ray
emitting sources (\eg a cluster of low-mass X-ray binaries, each
radiating at close to its Eddington luminosity, or a giant H {\small
II} region containing one or more X-ray binaries plus a diffuse
component) could not be spatially distinguished from {\it bona fide\/}
individual examples of the SES phenomenon.  This problem can now be
addressed with the order of magnitude improvement in spatial
resolution offered by {\it CHANDRA\/}.

A key to understanding the nature of the SES would be an
identification with a known class of object, and follow-up studies at
other wavelengths.  This is now also possible due to the high
astrometric accuracy achievable with \chan data ($\sim 0.6''$ RMS;
Proposer's Observatory Guide v3.0).  We may also address two
interesting and related questions.  The first is what is the effect of
an ionizing flux amounting to more than $10^{39} \rm ~erg~s^{-1}$ upon
its immediate surroundings?  The second concerns the local galactic
environment of the SES, and what does this tell us about the nature of
the source itself?

To these ends, we have recently acquired \chan ACIS-S X-ray data and
William Herschel Telescope/INTEGRAL field spectrograph optical data
for a number of prominent SES in the RW2000 sample.  In this letter we
report an important early result of this programme, the possible
identification of an optical counterpart to NGC 5204 X-1.  This SES is
located in the nearby ($d = 4.8$ Mpc; Tully 1988) magellanic-type
galaxy NGC 5204.  The presence of a luminous (log L$_{\rm X} \sim 39.8
\ergsec$) X-ray source in NGC 5204 was first shown in \ein data
(Fabbiano, Kim \& Trinchieri 1992).  This source has more recently
been reported in \ro HRI data, where it is seen to be slightly offset
from the nucleus of the galaxy (RW2000; Colbert \& Mushotzky 1999;
Lira, Lawrence \& Johnson 2000) and is therefore identified as a
possible SES as opposed to a low-luminosity active galactic nucleus
candidate.  In Section 2 we report the results of a 10 ks \chan X-ray
observation of this source.  The WHT/INTEGRAL follow-up observation is
then detailed in Section 3, followed (in Section 4) by a discussion of
what this reveals about the nature of the SES.

\section{X-ray data}

NGC 5204 X-1 was observed by \chan on 2001 January 9 for a single
exposure of 10.1 ks.  The target was positioned on the
back-illuminated ACIS-S S3 chip, which was operated in the standard
${1}\over{8}$ sub-array mode in order to alleviate the anticipated
effects of pile-up in the target source.  Data were reduced and
analysed using the {\small CIAO} software version 2.0.1, starting with
the level 2 event files.  The in-orbit background was at a constant
low-level during the observation, therefore we did not need to reject
any exposure time due to background flare contamination.

\begin{figure}
\centering
\includegraphics[width=6cm,angle=270]{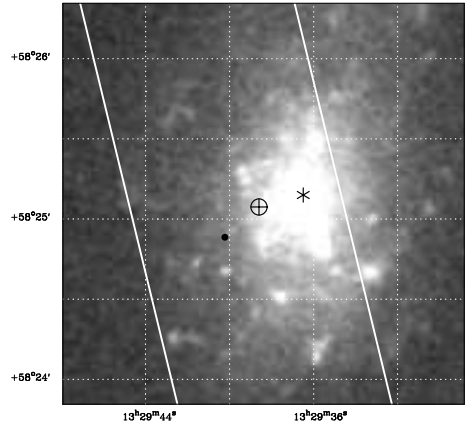}
\caption{{\it CHANDRA\/} source positions overlaid on to the Digitised
Sky Survey 2 image of the central $2.5' \times 2.5'$ region of NGC
5204.  The coverage of the {\it CHANDRA\/} sub-array is shown by the
parallel lines, and the galaxy nucleus is marked by an asterisk.  The
position of NGC 5204 X-1 is indicated by the crossed circle, and the
faint \chan detection by a small filled circle.}
\end{figure}

Figure 1 shows a Digitised Sky Survey 2 image of the galaxy.  Only two
X-ray sources are detected in the \chan data; NGC 5204 X-1 (at
$13^h29^m38.6^s +58^{\circ}25'06''$), and a very faint source which
lies $\sim 15''$ to its SE (at $13^h29^m40.3^s +58^{\circ}24'54''$).
The nucleus of the galaxy does not contain a luminous X-ray source.
NGC 5204 X-1 appears point-like.  Fitting the 2-D profile of the X-ray
source with a Gaussian model confirms that it is unresolved - the
half-energy width is found to be $0.8''$, which matches that found in
an on-axis ACIS observation of the point source PG1634-706 (\chan
Proposer's Observatory Guide, Figure 6.3).  We can therefore rule out
any significant extended component of size greater than $\sim 0.5''$
($\equiv 12$ parsecs in NGC 5204) contributing to the X-ray emission.

\begin{figure}
\centering
\includegraphics[width=4.5cm,angle=270]{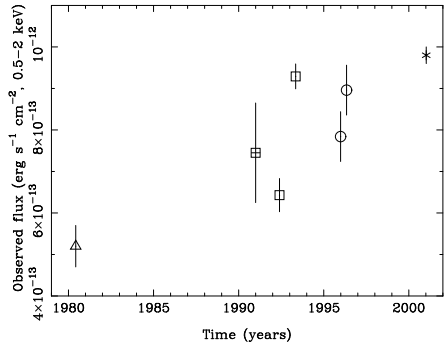}
\caption{Long-term X-ray lightcurve of NGC 5204 X-1.  This covers 7
observations over 20 years.  \ein data is indicated by a triangle, \ro
PSPC by squares, \ro HRI by circles and the new \chan data point by an
asterisk.  The count rates (converted to fluxes as per the text) were
taken from: Fabbiano, Kim \& Trinchieri (1992; \ein IPC); the \ro
all-sky survey bright source catalog (Voges et al. 1999) and WGACAT
(White, Giommi \& Angelini 1994) (\ro PSPC); RW2000 (\ro HRI).  Errors
in both plots are 1$\sigma$.}
\end{figure}

NGC 5204 X-1 has an observed ACIS-S count rate of 0.41 $\ctsec$ and so
the observation suffers from pile-up at the $\sim 10\%$ level.  This
leads to a small distortion of the X-ray spectrum towards a harder
form, and a slight underestimate of the true count rate; however, for
our present purpose we make a preliminary spectral analysis ignoring
pile-up effects (a more complete study will be the subject of future
work).  An X-ray spectrum was extracted for NGC 5204 X-1 using the
appropriate {\small CIAO} tools, and analysed using the standard X-ray
spectral fitting package {\small XSPEC}.  We limited spectral analysis
to the 0.5 - 10 keV band; however, in reality very few counts were
observed above 5 keV. The spectrum appears relatively featureless, and
the best fits to the data were found to be either a simple power-law
continuum absorbed by cold, neutral material (\nh $= 1.6 \pm 0.3
\times 10^{21} \atpcm$, $\Gamma =2.4 \pm 0.1$, $\chi^2_{\nu} = 1.27$
for 112 d.o.f.) or an absorbed thermal bremsstrahlung spectrum ( \nh
$= 5.4 \pm 1.7\times 10^{20} \atpcm$, $kT = 2.6 \pm 0.3$ keV,
$\chi^2_{\nu} = 1.25$).  Both a pure thermal plasma (MEKAL) model and
the ``multi-colour disc blackbody'' (MCD) model favoured in \asca
studies of ULX (\eg Makishima et al. 2000) were rejected.  More
complex models do not significantly improve the spectral fit.  It is
notable that the source spectrum appears soft, even after any
hardening due to pile-up.  The observed flux was $1.9 \times 10^{-12}
\ergcms$ (0.5 - 8 keV), which equates to a luminosity of $5.2 \times
10^{39} \ergsec$ assuming the source is located at the distance of NGC
5204.

The temporal characteristics of the source were also investigated.
The short-term lightcurve from the ACIS-S observation shows no
detectable variability over the 10ks observation (a $\chi^2$ test
against the hypothesis that the flux level remains constant gives
$\chi^2 = 31$ for 50 degrees of freedom).  However, long-term
variability is detected.  In Figure 2 we show the long-term
lightcurve, composed of archival measurements from the \ein and \ro
missions plus the new data point.  The lightcurve uses the best fit
model to the \chan data and, using either {\small PIMMS} or {\small
XSPEC}, folds it through the response of each mission which detected
NGC 5204 X-1.  In each case the flux was normalised to the observed
count rate, and fluxes are compared in the 0.5-2 keV band (chosen as
it is covered by all the relevant missions).  The resulting lightcurve
shows a doubling of the flux in the 20 years between the \ein IPC and
\chan ACIS-S observations, suggesting that the rising flux may
represent a long-term trend.  However, medium-term variability is also
apparent in the \ro PSPC observations, showing a flux increase of
$\sim 50\%$ in 11 months between 1993 and 1994.  The latter
variability is a strong pointer to the origin of the X-ray emission in
accretion onto a compact object.

\section{Optical data}

The optical data were taken on the nights of 2001 February 1 and 2,
using the INTEGRAL field spectrograph (Arribas et al. 1998) mounted on
the William Herschel Telescope, La Palma.  INTEGRAL provides
simultaneous optical spectroscopy and, through the relative positions
of the fibres, imaging.  We used the ``SB2'' fibre configuration,
which gave us a field-of-view of $16.5 \times 12.3$ arcsec$^2$ covered
by 189 fibres, each of diameter 0.9$''$.  In this configuration an
additional 30 fibres, at a radius of 45$''$, provide a measure of the
local sky background for each observation.  The R600R grating was
used, giving a wavelength coverage of $\sim 4500 - 7500$ \AA~ and
resolution $\sim 6$ \AA, and data were read out through WYFFOS.  The
target was observed for a total of 2 hours, split into three sections
over the two nights.

The fibre data were reduced in the standard fashion.  This will be
discussed in detail in a future paper (Goad et al., in preparation).
The data reduction ultimately provided a flux-calibrated spectrum for
each individual fibre, which were used to form the multi-colour narrow
band images.




\begin{figure*}
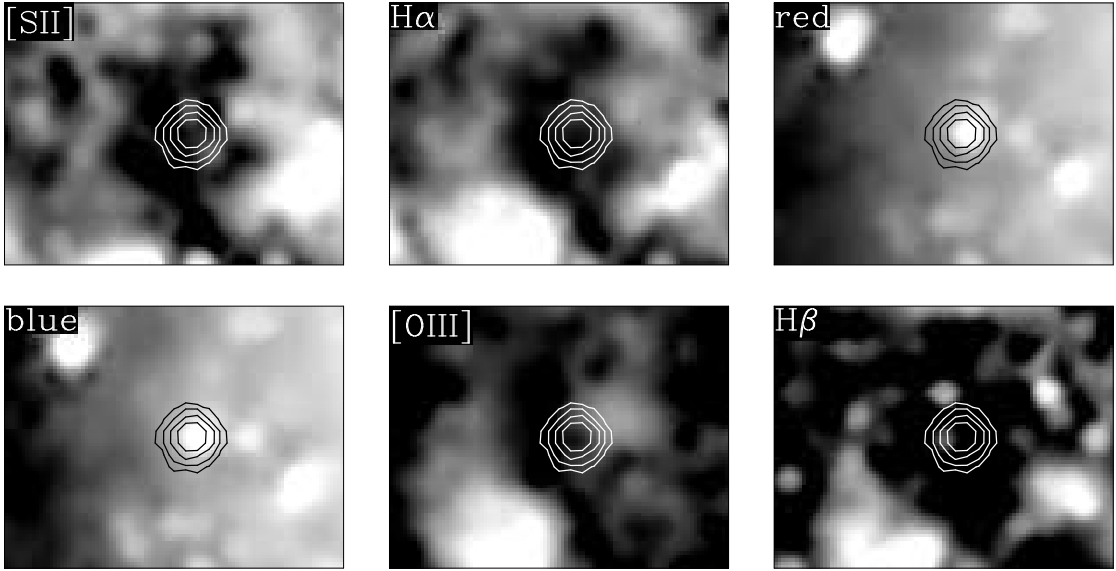

\centering
\includegraphics[width=3.5cm,angle=270]{n5204_int_overlay_s2.ps}\hspace*{0.5cm} 
\includegraphics[width=3.5cm,angle=270]{n5204_int_overlay_ha.ps}\hspace*{0.5cm}
\includegraphics[width=3.5cm,angle=270]{n5204_int_overlay_rcont.ps}\vspace*{0.5cm}
\includegraphics[width=3.5cm,angle=270]{n5204_int_overlay_bcont.ps}\hspace*{0.5cm}
\includegraphics[width=3.5cm,angle=270]{n5204_int_overlay_o3.ps}\hspace*{0.5cm}
\includegraphics[width=3.5cm,angle=270]{n5204_int_overlay_hb.ps}
\caption{Reconstructed narrow-band INTEGRAL images.  The images are in 
decreasing wavelength order, from top left to bottom right, and are in
an arbitrary logarithmic greyscale with white corresponding to the
highest emission values.  Note that all emission-line images are
continuum-subtracted, and we use the 6000 - 6200 \AA~ and 5100 - 5200
\AA~ ranges for the red and blue continuum images respectively.  X-ray
contours from the \chan data are overlaid.}
\end{figure*}

Figure 3 shows six reconstructed images of the field centred on the
position of NGC 5204 X-1, in a variety of narrow continuum bands and
continuum-subtracted emission-line bands, with the \chan X-ray
emission contours overlaid.  It is immediately obvious that an optical
continuum source is present at or close to the position of the X-ray
source.  This source is not detected in any continuum-subtracted
emission line image.  Inspection of the data shows that the continuum
emission is detected in a single fibre, numbered 113, implying that
the continuum source is point-like at the $0.9''$ resolution of
INTEGRAL.  This fibre is positioned immediately to the west of the
central fibre (number 110) of the field-of-view, but lies within the
error circle for the X-ray source position\footnote{The positional
error radius is a combination of the $\la 1$ arcsec \chan position
error and a similar error on the INTEGRAL position, which in practise
encompasses the central fibre plus its 6 immediate neighbours}, and so
the source is confirmed as a potential optical counterpart to the
X-ray source.

It is possible that this source could be an unrelated interloper at
the X-ray source position.  We have estimated the probability of it
being a background or foreground object, unrelated to NGC 5204, using
an APM source list for a blank $30 \times 30$ arcmin$^2$ field offset
to the NE of NGC 5204.  The observed magnitude of the counterpart is
$m_v = 19.7$; the chance of a background or foreground object of this
magnitude falling within the error circle is very small ($P \approx 3
\times 10^{-4}$).  Similarly, we can use the INTEGRAL data to estimate
the chance of a random coincidence with a comparably bright source in
the region of NGC 5204 containing the counterpart, which gives a
probability $P \approx 0.08$.  It is very likely then that the optical
counterpart is indeed associated with the X-ray source.

In Figure 4 we present the INTEGRAL optical spectrum of the source.
The source has a very obvious blue continuum spectrum, with little
evidence for emission lines (indeed, the shallow emission lines
present such as H$\alpha$ are likely to be due to the small amounts of
ionised gas seen in the environment of NGC 5204 X-1 in Figure 3).
Note that the presence of strong sky lines and low signal-to-noise
made the background subtraction and flat-fielding particularly
difficult at either end of the wavelength range, but despite this the
spectrum is essentially featureless.  We discuss the implications of
this in detail in the next section.

A second interesting feature of Figure 3 is an apparent ``hole'' in
the ionised gas, seen predominantly to the east of NGC 5204 X-1 in the
[SII], H$\alpha$ and [OIII] images, and also observed to extend to
the west in the H$\beta$ image.  The continuum images show no obvious
sign of this being due to absorption, for example in a dust lane.  It
is at least plausible therefore that this is a cavity in the ISM of
NGC 5204, apparently centred on the position of the X-ray source, with
a diameter of $\sim 100 - 200$ parsecs.

\section{The nature of the beast}

\subsection{A foreground or background source?}

In the previous section we showed that the optical counterpart is very
likely to be associated with the X-ray source.  However, in the
absence of strong features in the optical spectrum we cannot confirm
that the counterpart is associated with NGC 5204.  This means that we
must consider the merits of known classes of featureless blue
continuum foreground and background sources as potential
identifications.

A possible foreground identification, based on the very blue optical
spectrum, would be an isolated white dwarf in our galaxy.  However,
the X-ray properties are not consistent with such a white dwarf, which
would typically only display X-ray emission at energies up to a few
hundred eV as opposed to the several keV observed (\eg Marsh et
al. 1997).

In terms of a background object, the best candidate would appear to be
a BL Lac object.  The featureless blue optical and X-ray continua, its
point-like nature in both X-rays and optical, and the long-term X-ray
variability are all consistent with the known properties of BL Lacs.
In order to investigate this possibility further, given that BL
Lac objects are radio-loud AGN, we searched the NRAO VLA Sky
Survey (NVSS; Condon et al. 1998) for a point-like radio continuum
source at the position of NGC 5204 X-1.  No source is found at the
position of NGC 5204 X-1; however, an NVSS source of strength 16.3 mJy
is present $(26 \pm 14)''$ to its south-west.  The difference in
positions implies that the NVSS source is unlikely to be a radio
counterpart to NGC 5204 X-1 (the astrometry of the survey is good to
$\sim 1''$ compared to USNO optical positions), however this cannot be
ruled out given that the spatial resolution of the survey is only
$45''$.  In fact, we note that another candidate exists that may be
the source of the radio continuum emission, namely a supernova remnant
detected by Matonick \& Fesen (1997), NGC 5204 \#3, at a distance of
$(20 \pm 12)''$ from the NVSS detection.  At the very least the
presence of the radio source may limit the sensitivity with which we
can claim a non-detection at the position of NGC 5204 X-1, above the
typical survey limit of $\sim 2.5$ mJy.

To ascertain whether NGC 5204 X-1 is actually a BL Lac, we derived
$\alpha_{\rm ro}$ and $\alpha_{\rm ox}$ values as per
Laurent-Muehleisen et al. (1999), for the cases in which the NVSS
source is/is not the counterpart.  If it is, then $\alpha_{\rm ro} =
0.49$ and $\alpha_{\rm ox} = 1.1$, which by Figure 3 of
Laurent-Muehleisen et al. (1999) places NGC 5204 X-1 firmly in the
parameter space occupied by BL Lac objects.  If the NVSS source is not
a counterpart, and we assume an upper limit of 2.5 mJy for the NVSS
source flux, then we get $\alpha_{\rm ro} \leq 0.33$ which implies the
source has at best a low $\alpha_{\rm ro}$ value for a BL Lac.  A
further tightening of the constraints on radio continuum emission by a
factor of a few would clearly put NGC 5204 X-1 outside the parameter
space occupied by BL Lac objects, and so a high resolution radio
continuum follow-up of this source is essential.

\subsection{NGC 5204 X-1 as a SES}

A further possibility for the origin of the blue continuum optical
spectrum is the emission from one or more O stars\footnote{We note
that though O star spectra generally display absorption features,
these are low contrast features and we do not have the signal-to-noise
to distinguish them in the current optical spectrum.}.  The observed
magnitude ($m_v = 19.7$) is far too faint for a Galactic O star.
Instead, if the emission comes from a source in NGC 5204, then its
absolute magnitude is $M_v = -8.7$ (for a distance of 4.8 Mpc); this
corresponds to the emission of $\sim 8 - 20$ ``typical''
supergiant/giant O stars (Zombeck 1990) within a $\sim 12$ parsec
region in NGC 5204.  Though it seems unlikely that such a compact
cluster of O stars may exist in apparent isolation in the galaxy, it
is possible that we may be observing one or more unusually luminous O
stars.  If the optical counterpart is indeed stellar, then this would
appear to confirm that the X-ray emission is originating in a binary
system.

\begin{figure}
\centering
\includegraphics[width=5cm,angle=270]{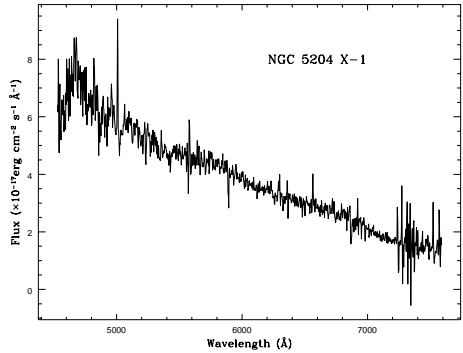}
\caption{INTEGRAL optical spectrum of the counterpart to NGC 5204 X-1.}
\end{figure}

Recent theoretical work (King et al. 2001) has discussed the likely
physical nature of the ultra-luminous compact X-ray sources (ULXs)
that constitute the accreting subset of SES.  They come to the
conclusion that, whilst accretion onto an intermediate-mass ($10^2 -
10^4$ M$_{\odot}$) black hole cannot be ruled out in individual cases,
the majority of ULXs may represent a short-lived but extremely common
stage in the evolution of many X-ray binaries, in which the X-ray
emission is mildly beamed.  The best candidate for this epoch is a
period of thermal-timescale mass transfer in intermediate- and
high-mass X-ray binaries.  The short lifetimes of high-mass X-ray
binaries and their association with star formation explains the
unusually high numbers of these objects present in very actively
star-forming galaxies (\eg `The Antennae'; Fabbiano, Zezas \& Murray
2001).  This explanation is fully consistent with our observation of
an O star companion to the accreting source, and may, if the mass
transfer is occuring due to the O star filling its Roche lobe as it
crosses the Hertzsprung gap, also explain in part the unusual optical
luminosity of the O star.

One final point to consider is the possible cavity in the ISM of NGC
5204 identified in the INTEGRAL emission-line images.  If this is a
true cavity centred on NGC 5204 X-1 then it is very plausible that NGC
5204 X-1 may be in some way responsible for this, possibly due to the
stellar wind from the O star (and its companions), or the progenitor
supernova of the compact source.  In either case, the presence of such
a cavity is further evidence that NGC 5204 X-1 lies within the
confines of the host galaxy.

\section{Conclusions}

We have used new \chan ACIS-S X-ray data and WHT/INTEGRAL integral
field spectroscopy to study the X-ray emission from the suspected SES
NGC 5204 X-1, and have identified a possible counterpart with a blue
continuum optical spectrum.  Though we cannot completely rule out the
scenario in which we are observing a BL Lac object behind NGC 5204,
the observational evidence is consistent with the scenario in which we
are viewing the beamed emission of a high-mass X-ray binary,
containing either a black hole or neutron star with an O star
companion, in a period of high mass transfer.

\vspace{0.2cm}

{\noindent \bf ACKNOWLEDGMENTS}

The authors would like to thank the INTEGRAL team for permitting us
access to unreleased reduction software.  We also thank Matt Burleigh
and Darach Watson for useful discussions.  TPR, MRG, PL and ADPH
gratefully acknowledge financial support from PPARC.  The Digitised
Sky Survey was produced at the Space Telescope Science Institute,
under US government grant NAG W-2166 from the original National
Geographic--Palomar Sky Survey plates.

\end{document}